\documentclass[12pt]{article}
\usepackage[english]{babel}
\usepackage[utf8]{inputenc}
\usepackage[T1]{fontenc}

\def\bPi{\mathbf{\Pi}}

\def\mK{\mathcal{K}}

\usepackage{amsmath}

\def\mC{\mathcal{C}}
\usepackage{amsfonts}

\usepackage{url}
\usepackage[dvips]{graphicx}
\usepackage{calc}
\usepackage{subfigure}
\usepackage{multirow}

\def\mL{\mathcal{L}}
\def\mH{\mathcal{H}}

\def\bx{\mathbf{x}}

\newcommand{\pb}[1]{\left\{#1\right\}}

\def\bA{\mathbf{A}}

\def\by{\mathbf{y}}
\def\bAi{\left(\bA^{-1}\right)}

\begin{document}
	\begin{titlepage}
	\begin{center}
		{\Large{ \bf Canonical Analysis of  Eddington Gravity}}
		
		\vspace{1em}  
		
		\vspace{1em} J. Kluso\v{n} 			
		\footnote{Email addresses:
			klu@physics.muni.cz (J.
			Kluso\v{n}) }\\
		\vspace{1em}
		\textit{Department of Theoretical Physics and
			Astrophysics, Faculty of Science,\\
			Masaryk University, Kotl\'a\v{r}sk\'a 2, 611 37, Brno, Czech Republic}
		
		\vskip 0.8cm
		
		%
		%
		%
		%
		%
		%
		
		\vskip 0.8cm
		
	\end{center}

\begin{abstract}
In this short note we perform canonical analysis of  Eddington gravity using L. Faddeev and Jackiw formalism.
We demonstrate that resulting canonical action has the same form as General Relativity canonical action 
which proves an equivalence of Eddington gravity and General Relativity in Hamiltonian formalism.

	\end{abstract}

\bigskip

\end{titlepage}

\newpage

\section{Introduction and Summary}\label{first}
In $1924$ Eddington proposed new intriguing formulation of gravity where
fundamental degrees of freedom is connection only
\cite{Eddington}. It turns out that this theory is equivalent to General Relativity 
in absence of matter in case of non-zero cosmological constant
\cite{Banados:2010ix}. 

On the other hand Eddington's gravity is incomplete since it is not well known how
it could be extended to couple to matter. 
 The most popular approach is in the extension of original Eddington form of gravity to so called Born-Infeld inspired gravity. First attempt for Born-Infeld gravity, not directly related to Eddington one,  was presented in  \cite{Deser:1998rj} where metric was fundamental degree of freedom however it was shown in the same paper that this formulation is plagued with ghost due to the fact that this action contains higher order derivatives of metric. Then it was recognized in
\cite{Vollick:2003qp,Vollick:2005gc,Banados:2010ix}
that   Born-Infeld gravity should be formulated in the first order formalism with  connection as dynamical variable 
which closely follows Eddington proposal. Another proposal how to include matter to Eddington formulation of gravity was presented in 
 \cite{Chakraborty:2020yag}. In this case the matter contribution is represented by inclusion of specific combination of matter stress energy tensor into action.  However drawback of  this construction is that the action implicitly depends on metric through  stress energy tensor. Another modification of Eddington gravity was suggested recently in 
\cite{Kluson:2025jpj} which is based on specific form of Born-Infeld inspired theory of gravity. It is clear that all these models modify equations of motion for gravity and hence they are not fully equivalent to Einstein equations of motion.

In this paper we focus on another unsolved problem of Eddington gravity which is its canonical formulation. It turns out that this is rather non-trivial problem since fundamental degrees of freedom are coefficients of connection $\Gamma^\rho_{\mu\nu}$. I is rather straightforward to determine conjugate momenta and find corresponding Hamiltonian. However it turns out that most of these conjugate momenta are zero which in the terminology of constraints systems developed by P. Dirac \cite{Dirac} means that they are primary constraints of theory. Then according to Dirac's procedure we should study stability of these constraints that generate additional secondary constraints and so on. It turns out that this analysis is rather complicated for gravity in the first order form as was demonstrated in \cite{McKeon:2010nf,Chishtie:2013fna,Kiriushcheva:2011aa,Kiriushcheva:2010ycc,Kiriushcheva:2010pia,Kiriushcheva:2006gp}. We rather use an alternative procedure pioneered by L. Faddeev and R. Jackiw
\cite{Faddeev:1988qp} (For review, see \cite{Jackiw:1993in}) which is equivalent to the standard Dirac approach as was shown in 
\cite{Garcia:1996ac}. In case of gravity similar procedure was used in  
\cite{Faddeev:1982id} and we proceed in similar way. Explicitly,  we  eliminate non-dynamical degrees of freedom by solving their equations of motion. Then we insert these solutions into canonical form of action and express it in terms of physical degrees of freedom. It turns out that even in this case lot of work has to be done in order to find true physical degrees of freedom. However in the end we obtain canonical action for Eddington gravity that has the same form as canonical action for general relativity which shows an equivalence of these two actions on canonical level. We mean that this is nice consistency check which is however preparation for more general situation when we include matter that can be performed in Born-Infeld like modification of General Relativity. 
This is very interesting goal for future project since we can expect that Hamiltonian will differ from General Relativity 
Hamiltonian. This problem is currently under investigation. 

Let us outline our results. We derived canonical form for Eddington gravity where the fundamental degree of freedom is variable \cite{McKeon:2010nf} $G^\lambda_{\mu\nu}=\Gamma^\lambda_{\mu\nu}-
\frac{1}{2}(\Gamma^\lambda_{\lambda \mu}\delta_\nu^\lambda+
\Gamma^\lambda_{\lambda\nu}\delta^\lambda_\mu)$ and their
conjugate momenta. Then we solved equations of motion for non-dynamical fields, inserted them into canonical form of action and identified physical degrees of freedom. Finally we showed that the canonical form of Eddington gravity is the same as canonical action for General Relativity that demonstrates  an equivalence between Eddington gravity and General Relativity in case of absence of matter. The natural extension of this analysis is find canonical formulation of Born-Infeld modified general relativity coupled to some simple form of matter where we can expect that resulting canonical form of action will differ from canonical action of General Relativity. This analysis is currently in progress.

\section{Canonical Formulation of  Eddington gravity}\label{second}
Eddington gravity is theory of gravity where the fundamental
degrees of freedom is connection only. In $d-$dimension 
 the action has the form 
\begin{equation}\label{act1}
	S=\int d^dx \sqrt{\det R_{(\mu\nu)}} \ , \quad 
R_{(\mu\nu)}=\frac{1}{2}(R_{\mu\nu}+R_{\nu\mu}) \ , 
\end{equation}
where Ricci tensor is defined as
\begin{equation}\label{defR}
	R_{\mu\nu}=\partial_\lambda\Gamma^\lambda_{\mu\nu}-
	\partial_\nu \Gamma^\lambda_{\lambda \mu}+\Gamma_{\mu\nu}^\lambda
	\Gamma^\sigma_{\sigma\lambda}-\Gamma^\lambda_{\sigma\mu}\Gamma^\sigma_{\lambda \nu} \ .
\end{equation}
In this paper we presume that coefficients of connections are 
symmetric 
 $\Gamma^\lambda_{\mu\nu}=\Gamma^\lambda_{\nu\mu}$ while   Ricci tensor is not symmetric due to the presence of second term in its definition (\ref{defR}). Since $\Gamma^\lambda_{\mu\nu}$ defines  connection they have
physical dimension $[M]$ for some mass scale and hence $R_{\mu\nu}$ has physical dimension $[M^2]$. As a result $\det R_{\mu\nu}$ has physical dimension $[M^{2d}]$ and  since physical dimension of $d^dx$ is $[M^{-d}]$ we get that  the action 
(\ref{act1}) is dimensionless. 

For further purposes we introduce new variable $G^\lambda_{\mu\nu}$ defined as 
\cite{Horava:1990ba}
\begin{equation}
	G_{\mu\nu}^\lambda=
	\Gamma_{\mu\nu}^\lambda-\frac{1}{2}(\delta_\mu^\lambda
	\Gamma^\rho_{\rho\nu}+\delta_\nu^\lambda \Gamma^\rho_{\rho\mu})
\end{equation}
so that $R_{(\mu\nu)}$ can be written as  
\begin{equation}\label{Rsim}
	R_{(\mu\nu)}=
	\partial_\lambda G^\lambda_{\mu\nu}+
	\frac{1}{d-1}G^\lambda_{\lambda\mu}G^\sigma_{\sigma\nu}
	-G^\lambda_{\sigma\mu}G^\sigma_{\lambda\nu}	 \ . 
\end{equation}
As a check note
\begin{eqnarray}
	\partial_\lambda G^\lambda_{\mu\nu}=
	\partial_\lambda \Gamma^\lambda_{\mu\nu}-
	\frac{1}{2}(\partial_\mu \Gamma^\rho_{\rho\nu}
	+\partial_\nu \Gamma^\rho_{\rho\mu}) \ , \quad 
	G^\lambda_{\lambda\mu}=\frac{1-d}{2}\Gamma^\lambda_{\lambda\mu}
	\nonumber \\
	G^\lambda_{\sigma\mu}G^\sigma_{\lambda\nu}=
	\Gamma^\lambda_{\sigma\mu}\Gamma^\sigma_{\lambda\nu}-\Gamma^\lambda_{\lambda\mu}
	\Gamma^\rho_{\rho\nu}-\Gamma^\lambda_{\mu\nu}\Gamma^\rho_{\rho\lambda}+
	\frac{1}{4}(d+3)\Gamma^\rho_{\rho\mu}\Gamma^\sigma_{\sigma\nu}
	\nonumber \\
\end{eqnarray}
and hence collecting all these terms together we obtain (\ref{Rsim}).

As the first step in the canonical analysis we determine conjugate momenta to 
 $G^\lambda_{\mu\nu}$
 from the action (\ref{act1})  
\begin{eqnarray}\label{Pimunu}
&&	\Pi^{\mu\nu}_0=\frac{\partial \mL}{\partial (\partial_0 G ^0_{\mu\nu})}=
	\frac{1}{2}\sqrt{\det \bA}\bAi^{\mu\nu}\equiv \bPi^{\mu\nu} \  ,  \nonumber \\
&&	\Pi^{\mu\nu}_i=\frac{\partial \mL}{\partial (\partial_0 G^i_{\mu\nu})}=0 \ , 
	\nonumber \\
\end{eqnarray}
where $\bAi$ is matrix inverse to $\bA_{\mu\nu}\equiv R_{(\mu\nu)}$
\begin{equation}
	\bA_{\mu\nu}\bAi^{\nu\rho}=\delta_\mu^\rho \ . 
\end{equation}
Clearly it is natural to demand that $\bA_{\mu\nu}$ is non-singular since in opposite case we would get vanishing 
action. With this natural presumption we obtain Hamiltonian density in the form
\begin{eqnarray}
&&	\mH=\bPi^{\mu\nu}\partial_0 G^0_{\mu\nu}-\mL
	=\bPi^{\mu\nu}\bA_{\mu\nu}-\nonumber \\
&&	-\bPi^{\mu\nu}(\partial_i G^i_{\mu\nu}+
	\frac{1}{d-1}G^\lambda_{\lambda\mu}G^\sigma_{\sigma\nu}-G_{\sigma\mu}^\lambda 
	G_{\lambda\nu}^\sigma)-\mL=
	\nonumber \\
&&	=\frac{d-2}{2}\sqrt{\det \bA}
	-\bPi^{\mu\nu}(\partial_i G^i_{\mu\nu}+
	\frac{1}{d-1}G^\lambda_{\lambda\mu}G^\sigma_{\sigma\nu}-G_{\sigma\mu}^\lambda 
	G_{\lambda\nu}^\sigma) \ . \nonumber \\
\end{eqnarray}
Finally from definition of $\bPi^{\mu\nu}$ we get 
\begin{eqnarray}
	\mathbf{\Pi}\equiv 	\det \bPi^{\mu\nu}=
	2^{-d}\det\bA^{\frac{d-2}{2}}
	\nonumber \\	
\end{eqnarray}
and hence
\begin{equation}
	\sqrt{\det \bA}=2^{\frac{d}{d-2}}\sqrt{\Pi^{\frac{2}{d-2}}} \ . 
\end{equation}
Using this result we obtain Hamiltonian density as function of canonical variables
\begin{eqnarray}\label{mHfinal}
	\mH=(d-2)2^{\frac{2}{d-2}}\sqrt{\bPi^{\frac{2}{d-2}}}
	-\bPi^{\mu\nu}(\partial_i G^i_{\mu\nu}+
	\frac{1}{d-1}G^\lambda_{\lambda\mu}G^\sigma_{\sigma\nu}-G_{\sigma\mu}^\lambda 
	G_{\lambda\nu}^\sigma)
	\nonumber \\
\end{eqnarray}
so that canonical form of the action is equal to 
\begin{equation}\label{actcan}
	S=\int d^d x (\bPi^{\mu\nu}\partial_0 G_{\mu\nu}^0-\mH) \ . 
\end{equation}
From (\ref{mHfinal}) we see that  $G_{\mu\nu}^i$ is non-dynamical which is a consequence of the fact that $\Pi^{\mu\nu}_i=0$.  As the next step we have two options. The first one is 
 standart Dirac-Bergmann analysis of constraints systems when we say that $\Pi^{\mu\nu}_i=0$ are primary constraints of theory and require their preservation during  time evolution of system. This procedure could potentially lead to secondary constraints when we should again check their stability and so on. In the similar way  first order formulation of gravity was studied in  
\cite{McKeon:2010nf,Chishtie:2013fna,Kiriushcheva:2011aa,Kiriushcheva:2010ycc,Kiriushcheva:2010pia,Kiriushcheva:2006gp}. However such an analysis is rather awkward since during this procedure many new additional constraints are generated and  we have to check their stability and determine which constraints are first class and which second class and so on.

 For that reason we rather follow alternative approach based on work L. Faddeev and R. Jackiw 
\cite{Faddeev:1988qp} (For review, see \cite{Jackiw:1993in}) which is equivalent to the standard Dirac approach as was shown in 
\cite{Garcia:1996ac}. In case of gravity similar procedure was used in  
\cite{Faddeev:1982id}. The basic idea is to  eliminate non-dynamical degrees of freedom by solving their equations of motion. In more details, variation of the canonical action 
(\ref{actcan}) with respect to  $G_{\mu\nu}^i$ gives 
\begin{eqnarray}
	-\partial_i\bPi^{\mu\nu}+\frac{1}{d-1}
	(\bPi^{\nu\beta}\delta_i^\mu+\bPi^{\mu\beta}\delta_i^\nu)
	G^\sigma_{\sigma\beta}-G^\mu_{i\beta}\bPi^{\beta\nu}-
	G^\nu_{i\beta}\bPi^{\beta\mu}\equiv -\Sigma^{\mu\nu}_i=0\nonumber \\	
\end{eqnarray}
that can be rewritten into more transparent form when we replace $G^\lambda_{\mu\nu}$ with $\Gamma^\lambda_{\mu\nu}$
\begin{eqnarray}\label{defSiamaimunu}
&&	\Sigma_i^{\mu\nu}=
\partial_i\bPi^{\mu\nu}+\Gamma^{\mu}_{i\beta}\bPi^{\beta\nu}+
	\Gamma^\nu_{i\beta}\bPi^{\beta\mu}-\Gamma^\sigma_{\sigma i}
	\bPi^{\mu\nu}= 0 \nonumber \\
\end{eqnarray}
which has the form of covariant derivative of $\bPi^{\mu\nu}$ where the last term in (\ref{defSiamaimunu}) is a consequence of the fact that 
$\bPi^{\mu\nu}$ is tensor density. 
Thanks to these equations we can express 
non-propagating degrees of freedom using the canonical ones. 

Firstly we start with the equation
\begin{eqnarray}
	\Sigma^{00}_i=\partial_i\bPi^{00}+2\Gamma^0_{i0}\bPi^{0 0}
+2\Gamma^0_{im}\bPi^{m0}	
	-\Gamma^m_{m i}\bPi^{00}-\Gamma^0_{0i}\bPi^{00}=0
\end{eqnarray}
that ca be solved for $\Gamma^0_{0i}$ as 
\begin{equation}
\Gamma^0_{i0}-\frac{1}{\bPi^{00}}(\nabla_i \bPi^{00}+\Gamma^0_{im}\bPi^{m0}) \ , \nonumber \\ 
\end{equation}
where we introduced spatial covariant derivative 
\begin{equation}\label{covderPioo}
\nabla_i\bPi^{00}=\partial_i \bPi^{00}-\gamma^m_{mi}\bPi^{00} \ ,  
\end{equation}
with coefficients of connection given as 
\begin{equation}
	\gamma^i_{jk}=\Gamma^i_{jk}-\frac{\bPi^{i0}}{\bPi^{00}}\Gamma^0_{jk} \ . 
\end{equation}
Note that the second term in (\ref{covderPioo}) is necessary since $\bPi^{00}$ is scalar density.
Further, solving equations $\Sigma^{0i}_j=0$ we obtain 
\begin{eqnarray}
\Gamma^j_{0i}=-\frac{1}{\bPi^{00}}
(\nabla_i\bPi^{0j}+\Gamma^0_{im}\bPi^{mj}) \ , \nonumber \\
\end{eqnarray}
where
\begin{equation}
	\nabla_i\bPi^{0j}=\partial_i\bPi^{0j}+\gamma^j_{ik}\bPi^{k0}-
	\gamma^m_{mi}\bPi^{0j} \ . 
\end{equation}
Using these results we can now manipulate with kinetic term in the canonical action and we obtain (after integration by parts when we ignore boundary terms)
\begin{eqnarray}\label{kinterm}
&&	\int d^dx (\bPi^{\mu\nu}\partial_0 G^0_{\mu\nu})=
	-\int d^dx \partial_0\bPi^{\mu\nu}G_{\mu\nu}^0=
\nonumber \\
&&=\int d^dx\left
(\partial_0(\bPi^{0i}\bPi^{0j}-\bPi^{00}\bPi^{ij})\frac{1}{\bPi^{00}}G^0_{ij}-\frac{\partial_0\bPi^{00}}{\bPi^{00}}\partial_i\bPi^{0i}+
\frac{\partial_i\bPi^{00}}{\bPi^{00}}\partial_0\bPi^{0i}\right) \nonumber \\
&&	=\int d^dx \partial_0 q^{ij}\mK_{ij}
+\int d^dx \left(-\frac{\partial_0\bPi^{00}}{\bPi^{00}}\partial_i\bPi^{0i}+
\frac{\partial_i\bPi^{00}}{\bPi^{00}}\partial_0\bPi^{0i}\right)
 \ , 
\end{eqnarray}
where in the last step 
 we introduced the matrix $q^{ij}$ and its conjugate momentum $\mK_{ij}$ defined as 
\begin{equation}
	q^{ij}=\bPi^{00}\bPi^{ij}-
\bPi^{0i}\bPi^{0j} \ , \quad 
\mK_{ij}=-\frac{1}{\bPi^{00}}\Gamma^0_{ij} \ .
\end{equation} 
As the final step let us write last two terms in  (\ref{kinterm}) as
\begin{eqnarray}
&&	\int d^dx (-\frac{\partial_0\bPi^{00}}{\bPi^{00}}\partial_i\bPi^{0i}+
	\frac{\partial_i\bPi^{00}}{\bPi^{00}}\partial_0\bPi^{0i})=
	\nonumber \\
&&	=\int d^dx  (-\partial_0 (\ln \bPi^{00}\partial_i\bPi^{0i})+
	\partial_i (\ln \bPi^{00}\partial_0\bPi^{0i}))
	\nonumber \\
\end{eqnarray}
that gives zero contribution to the action when we perform integration by parts and ignore boundary terms.  Returning to 
(\ref{kinterm}) we see that there is  natural simplectic structure between $q^{ij}$ and $\mK_{kl}$.  In fact, if we start with original Poisson brackets for $G^\rho_{\mu\nu}$ and $\Pi^\sigma_{\alpha\beta}$ in the form 
\begin{equation}
	\pb{G^\rho_{\mu\nu}(\bx),\Pi_\sigma^{\alpha\beta}(\by)}=
\frac{1}{2}\delta_\sigma^\rho(\delta_\mu^\alpha\delta_\nu^\beta+\delta_\mu^\beta\delta_\nu^\alpha)\delta(\bx-\by)
\end{equation}
we find that $q^{ij}$ and $\mK_{mn}$ have following Poisson 
brackets
\begin{equation}\label{pbqk}
	\pb{q^{ij}(\bx),\mK_{kl}(\by)}=
	-\pb{\bPi^{00}(\bx)\Pi^{ij}_0(\bx),\frac{1}{\bPi^{00}(\by)}
		G^0_{kl}(\by)}=\frac{1}{2}(\delta^i_k\delta^j_l+\delta^i_l\delta^j_k)
	\delta(\bx-\by) \ . 
\end{equation}
There is another reason why  $q^{ij}$ is fundamental physical variable of Eddington gravity. Let us consider
 linear combinations of equations of motion $\Sigma^{\mu\nu}_i=0$ 
\begin{eqnarray}
&&	\bPi^{00}\Sigma_k^{ij}+\bPi^{ij}\Sigma_k^{00}-
	\bPi^{0i}\Sigma_k^{0j}-\bPi^{0j}\Sigma^{0i}_k=
	\nonumber \\
&&=\partial_k q^{ij}+(\Gamma^i_{km}-\frac{\bPi^{i0}}{\bPi^{00}}\Gamma^j_{km})q^{mj}+(\Gamma^j_{km}-\frac{\bPi^{j0}}{\bPi^{00}}\Gamma^0_{km})q^{mi}
-2(\Gamma^m_{mk}-\frac{\bPi^{m0}}{\bPi^{00}}\Gamma^m_{mk})q^{ij}=0\nonumber \\
\end{eqnarray}	
that can be rewritten into the form 
\begin{eqnarray}\label{partq}
	\partial_k q^{ij}+\gamma^i_{km}q^{mj}+\gamma^j_{km}q^{mi}
	-2\gamma^m_{mk}q^{ij}=0  \nonumber \\
\end{eqnarray}
This has the form of compatibility condition for variable $q^{ij}$ and covariant derivative
defined with connection $\gamma^k_{ij}$ if we take into account that 
$q^{ij}$ is tensor density of rank $2$. Note that the 
matrix inverse to $q^{ij}$ is equal to 
\begin{equation}
	q_{ij}=\frac{1}{\bPi^{00}}\bPi_{ij} \ , \quad  q^{ij}q_{jk}=\delta^i_k \ , 
\end{equation}
where $\bPi_{ij}$ are spatial components of the matrix $\bPi_{\mu\nu}$ which is inverse to $
\bPi^{\mu\nu}$.
Note also that (\ref{partq}) can be rewritten into the form 
\begin{eqnarray}
\partial_k q_{ij}-\gamma^m_{ki}q_{mj}-\gamma^m_{kj}q_{mi}+
2\gamma^n_{nk}q_{pm}=0 \ . \nonumber \\
\end{eqnarray}
As we argued above  natural 
dynamical variables are $q^{ij}$ and $\mK_{ij}$ and hence
we should express $\mH$ as function of these variables only.
To do this we firstly express Hamiltonian density 
as function of $\Gamma$ 
\begin{eqnarray}
\mH
=(d-2)2^{\frac{2}{d-2}}\sqrt{\bPi^{\frac{2}{d-2}}}-\bPi^{\mu\nu}\partial_i\Gamma^i_{\mu\nu}+
\bPi^{i\mu}\partial_i \Gamma^\rho_{\rho\mu}+
\bPi^{\mu\nu}	\Gamma^\lambda_{\sigma\mu}\Gamma^\sigma_{\lambda\nu}-\bPi^{\mu\nu}\Gamma^\lambda_{\mu\nu}\Gamma^\rho_{\rho\lambda} \nonumber \\
\end{eqnarray}
that, after integration by parts can be explicitly written as 
\begin{eqnarray}\label{mHsplit}
&&	\mH=(d-2)2^{\frac{2}{d-2}}\sqrt{\bPi^{\frac{2}{d-2}}}
	-\Gamma^0_{00}\Sigma_m^{0m}+
	\Gamma^i_{00}\Sigma_i^{00}
+\nonumber \\
&&+\partial_i\bPi^{mn}\Gamma^i_{mn}+2\partial_i\bPi^{0m}\Gamma_{0m}^i
-\partial_i\bPi^{im}\Gamma^0_{0m}
-\partial_i\bPi^{i0}\Gamma^m_{m0}-\partial_i\bPi^{im}\Gamma^n_{nm}
\nonumber \\
&&+\bPi^{00}\Gamma^m_{i0}\Gamma^i_{m0}+2\bPi^{0i}(\Gamma^n_{m0}\Gamma^m_{ni}
-\Gamma^0_{0i}\Gamma^m_{m0}-\Gamma^m_{0i}\Gamma^n_{nm})+
\nonumber \\
&&+\bPi^{ij}(\Gamma^0_{0i}\Gamma^0_{0j}+\Gamma^m_{0i}\Gamma^0_{mj}+
\Gamma^0_{mi}\Gamma^m_{0j}+\Gamma^n_{mi}\Gamma^m_{nj}-\Gamma^0_{ij}\Gamma^m_{m0}-
\Gamma^m_{ij}\Gamma^0_{0m}-\Gamma^m_{ij}\Gamma^n_{nm}) \ .
\nonumber \\
\end{eqnarray}
We see that $\Gamma^\mu_{00}$ appear in the Hamiltonian  as Lagrange multipliers for equations of motion $\Sigma^{0m}_m$ and $\Sigma^{00}_i$ so that they decouple from the theory. 
As the next step we replace $\Gamma^i_{jk},\Gamma^j_{0i}$ and 
$\Gamma^0_{0i}$ in the Hamiltonian (\ref{mHsplit})
with following relations 
\begin{equation}\label{Gammagamma}
	\Gamma^i_{jk}=\gamma^i_{jk}-\bPi^{i0}\mK_{jk}
\end{equation}
and 
\begin{equation}\label{Gamma0ij}
	\Gamma^j_{0i}=-\frac{1}{\bPi^{00}}\nabla_i\bPi^{0j}+\mK_{im}\bPi^{mj} \ ,  \quad 
	\Gamma^0_{0i}=-\frac{1}{\bPi^{00}}\nabla_i\bPi^{00}+\mK_{im}\bPi^{m0} \ . 
\end{equation}
It is clear from (\ref{mHsplit}) that generally inserting 
(\ref{Gammagamma}) and (\ref{Gamma0ij}) into it gives very complicated expression. Then it is convenient to collect terms of order $\mK^2,\mK^1$ and $\mK^0$ and we find that Hamiltonian simplifies considerably.

The simplest situation occurs with terms that are  
quadratic in $\mK$ and  we obtain that their contribution is equal to
\begin{eqnarray}\label{quadK}
\frac{1}{\bPi^{00}}\mK_{ij}q^{im}q^{jn}\mK_{mn}-
\frac{1}{\bPi^{00}}\mK_{ij}q^{ij}\mK_{mn}q^{mn} \ . 
\nonumber \\
\end{eqnarray}
In case of  terms  linear in $\mK$ the calculations are more involved but we again obtain simple result 
\begin{equation}\label{linK}
2\nabla_i(\frac{\bPi^{0m}}{\bPi^{00}})q^{in}\mK_{mn} 
	-2\nabla_m(\frac{\bPi^{0m}}{\bPi^{00}})q^{mn}\mK_{mn}
\end{equation}
Finally we collect all  terms of zero order in $\mK$ that appear in (\ref{mHsplit}) and we get 
\begin{eqnarray}
&&	-2\frac{\bPi^{0i}}{\bPi^{00}}\nabla_m\bPi^{0n}\gamma^m_{ni}
	-2\bPi^{0i}\frac{\nabla_i\bPi^{00}}{(\bPi^{00})^2}\nabla_m\bPi^{m0}+
	2\frac{\bPi^{0i}}{\bPi^{00}}\nabla_i\bPi^{0m}\gamma^n_{nm}+\nonumber \\
&&+	\partial_i\bPi^{mn}\gamma^i_{mn}-\partial_i\bPi^{im}\gamma^n_{nm}
	-2\partial_i\bPi^{0m}\frac{\nabla_m\bPi^{0i}}{\bPi^{00}}+
	\partial_i\bPi^{im}\frac{\nabla_m \bPi^{00}}{\bPi^{00}}
	+\partial_i\bPi^{i0}\frac{\nabla_m\bPi^{0m}}{\bPi^{00}}-
	\nonumber \\	
&&	+\frac{1}{\bPi^{00}}\nabla_m\bPi^{0n}\nabla_n\bPi^{0m}
+\frac{1}{(\bPi^{00})^2}\nabla_i\bPi^{00}\bPi^{ij}\nabla_j\bPi^{00}+\gamma^m_{ij}\bPi^{ij}\frac{\nabla_m\bPi^{00}}{\bPi^{00}}
+\nonumber \\
&&+\gamma^n_{mi}\bPi^{ij}\gamma^m_{jn}-\gamma^m_{ij}\bPi^{ij}
\gamma^n_{nm}=
\nonumber \\	
&&=\frac{1}{\bPi^{00}} q^{nm}\partial_n\gamma^p_{pm}-
\frac{1}{\bPi^{00}} q^{ij}\partial_m\gamma^m_{ij}+
\frac{1}{\bPi^{00}}\gamma^n_{mi}q^{ij}\gamma^m_{nj}-
\frac{1}{\bPi^{00}}\gamma^m_{ij}q^{ij}\gamma^p_{pm}-
\nonumber \\
&&-\partial_i(\frac{\bPi^{0j}}{\bPi^{00}})\partial_j\bPi^{0i}+
\partial_i(\frac{\bPi^{0i}}{\bPi^{00}})\partial_m\bPi^{0m} \ . 
\nonumber \\
\end{eqnarray}
Note that  last two terms in the expression above can be written as 
\begin{eqnarray}
	-\partial_i(\frac{\bPi^{0j}}{\bPi^{00}})\partial_j\bPi^{0i}+
	\partial_i(\frac{\bPi^{0i}}{\bPi^{00}})\partial_m\bPi^{0m}
	\nonumber \\
=	-\partial_i(\frac{\bPi^{0j}}{\bPi^{00}}\partial_j\bPi^{0i})+
	\partial_i(\frac{\bPi^{0i}}{\bPi^{00}}\partial_m\bPi^{0m})
	\nonumber \\
\end{eqnarray}
so that can be ignored when we neglect boundary terms in the action.
Collecting all these terms together we obtain canonical form of the action 
\begin{eqnarray}\label{actcan1}
	S=\int d^dx \left(\partial_t q^{ij}\mK_{ij}-\frac{1}{\bPi^{00}}\mC-\frac{\bPi^{0i}}{\bPi^{00}}\mC_i\right) \ , \nonumber \\
\end{eqnarray}
where we performed integration by parts and where $\mC$ and $\mC_i$ are defined as
\begin{eqnarray}
&&	\mC=(d-2)2^{\frac{2}{d-2}}(\det q^{ij})^{\frac{1}{d-2}}+\mK_{ij}q^{im}q^{jn}\mK_{mn}-
	\mK_{ij}q^{ij}\mK_{mn}q^{mn}+\nonumber \\
&&	+ q^{nm}\partial_n\gamma^p_{pm}-
	 q^{ij}\partial_m\gamma^m_{ij}+
	\gamma^n_{mi}q^{ij}\gamma^m_{nj}-
	\gamma^m_{ij}q^{ij}\gamma^p_{pm} \ , 
 \nonumber \\
&&	\mC_i=-2\nabla_m (q^{mn}\mK_{in})+2\nabla_i(q^{mn}\mK_{mn}) \ . \  \nonumber \\
\end{eqnarray}
We see that the propagating degrees of freedom are $q^{ij}$ and their conjugate momenta $\mK_{ij}$ while $\bPi^{00},\bPi^{0i}$ have interpretation as Lagrange multipliers so that 
the action (\ref{actcan1}) has similar structure as canonical 
action for General Relativity \cite{Dirac:1958sc,Arnowitt:1962hi}. To see this in more detail
it is instructive to define matrix $h_{ij}$ that is related to $q_{ij}$ by following relation
\begin{equation}\label{defq}
	q_{ij}=\sqrt{q}^p h_{ij} \ , 
\end{equation}
where $p$ is unknown number that will be determined from compatibility condition for $q_{ij}$ that, using
(\ref{defq}), gives 
\begin{equation}
	p\frac{\partial_k\sqrt{q}}{\sqrt{q}}+\sqrt{q}^p\partial_k h_{ij}-\gamma^m_{ki}\sqrt{q}^p h_{mj}
	-\gamma^m_{kj}\sqrt{q}^p h_{mi}+\frac{2}{d-2}\frac{\partial_k\sqrt{q}}{\sqrt{q}}=0
\end{equation}
and if we demand that terms proportional to $\partial_k\sqrt{q}$ cancel we obtain 
\begin{equation}
	p=-\frac{2}{d-2} \ . 
\end{equation}
As a result we get that 
  the matrix $h_{ij}$ obeys the compatibility condition 
\begin{equation}
	\partial_k h_{ij}-\gamma^m_{ki}h_{mj}-\gamma^m_{kj}h_{mi}=0 \ . 
\end{equation}
In other words $\gamma^i_{jk}$ are uniquely determined by $h_{ij}$ as 
\begin{equation}
	\gamma^i_{jk}=\frac{1}{2}h^{im}(\partial_j h_{mk}+\partial_k h_{mj}-\partial_m h_{jk}) \ , 
\end{equation}
where the matrix $h^{ij}$ is inverse to $h_{ij}$ 
so that $h_{ij}h^{jk}=\delta_i^k$.
Further, using (\ref{defq}) we obtain 
\begin{equation}
	\partial_0 q^{ij}
	=-\det h h^{ik}(\partial_0 h_{kl}-h_{kl}\partial_0 h_{mn}h^{mn})h^{lj}
\end{equation}
and hence we can write kinetic term in the action as 
\begin{eqnarray}
\int d^dx \partial_0 q^{ij}\mK_{ij}=
\int d^dx \partial_0 h_{ij}\frac{\det h}{\bPi^{00}}(h^{ik}\Gamma_{kl}^0h^{lj}-h^{ij}
h^{mn}\Gamma^0_{mn}) \  \nonumber \\
\end{eqnarray}
so that it is natural to identify  momentum $\pi^{ij}$ conjugate to $h_{ij}$ as
\begin{equation}\label{piij}
\pi^{ij}=\det h(- h^{im}\mK_{mn}h^{nj}+h^{ij}h^{mn}\mK_{mn})
\end{equation}
so that we have 
 an inverse relation 
\begin{equation}
	\mK_{ij}=-\frac{1}{\det h} h_{im}(\pi^{mn}-\frac{\pi}{d-2}h^{mn})h_{nj} \ , \quad 
	\mK\equiv \mK_{ij}h^{ij}=\frac{\pi}{\det h(d-2)} \ , \quad  \pi\equiv \pi^{ij}h_{ij} \ . 
\end{equation}
Then the quadratic term in $\mC$ can be written as  
\begin{eqnarray}
&&	\frac{1}{\bPi^{00}}
\mK_{ij}q^{im}q^{jn}\mK_{mn}-\frac{1}{\bPi^{00}}\mK_{ij}q^{ij}\mK_{mn}q^{mn}=
\nonumber \\
&&=\frac{1}{\bPi^{00}}(\pi^{ij}q_{im}q_{jn}\pi^{mn}-\frac{1}{d-2}\pi^2)  \ . 
\nonumber \\
\end{eqnarray}
Finally term linear in $\mK$ is equal to 
\begin{eqnarray}
	2\nabla_i(\frac{\bPi^{0m}}{\bPi^{00}})q^{in}\mK_{mn}-
	2\nabla_m(\frac{\bPi^{0m}}{\bPi^{00}})q^{mn}\mK_{mn}=
-2\nabla_i(\frac{\bPi^{0m}}{\bPi^{00}})h_{mn}\pi^{ni} \ . 
\nonumber \\
\end{eqnarray}
 Note also that $\bPi^{00}$ and $\bPi^{0i}$ are tensor densities so that it is natural to write them as
\begin{equation}
	\bPi^{00}=\sqrt{h}\frac{1}{\lambda} \ , \bPi^{0i}=\sqrt{h}\frac{\lambda^i}{ \lambda}\ , 
\end{equation}
where $\lambda$ and $\lambda^i$ are scalar functions. Then we obtain canonical action in the form 
\begin{equation}
	S=\int d^dx (\partial_t h^{ij}\pi_{ij}-\lambda \tilde{\mC}-\lambda^i\tilde{\mC}_i) \ , 
\end{equation}
where
\begin{eqnarray}
	\tilde{\mC}=(d-2)2^{\frac{2}{d-2}}\sqrt{h}+\frac{1}{\sqrt{h}}(\pi^{ij}h_{im}h_{jn}\pi^{mn}-\frac{1}{d-2}
	\pi^2)-\sqrt{h}{}^{(3)}R \ , \nonumber \\
{}^{(3)}R= h^{ij}(-\partial_i\gamma^p_{pj}+
\partial_m\gamma^m_{ij}-
\gamma^n_{mi}\gamma^m_{nj}+
\gamma^m_{ij}\gamma^p_{pm}) \ , \quad 
	\mC_i=-2\nabla_k (h_{ij}\pi^{jk}) \ . 
\end{eqnarray}
This is the final form of the canonical action for Eddington gravity which has the same form as canonical action for 
General Relativity \cite{Dirac:1958sc,Arnowitt:1962hi} that
demonstrates  an equivalence between Eddington gravity and General Relativity at the canonical formalism. It is also interesting that  components of metric arise as conjugate momenta
 to $G^0_{\mu\nu}$. 

Finally it is instructive to determine Poisson brackets between $h_{ij}$ and $\pi^{kl}$. Using
(\ref{defq}) and (\ref{piij}) and  with the help of Poisson bracket between 
$q^{ij}$ and $\mK_{kl}$ given in (\ref{pbqk}) we obtain
\begin{eqnarray}
	\pb{h_{ij}(\bx),\pi^{kl}(\by)}
	=\frac{1}{2}(\delta_i^k\delta_j^l+\delta_i^l\delta_j^k)\delta(\bx-\by) \  \nonumber \\
\end{eqnarray}
which is again nice consistency check. 

{\bf Acknowledgment:}

This work  is supported by the grant “Dualitites and higher order derivatives” (GA23-06498S) from the Czech Science Foundation (GACR).

\end{document}